\documentclass[11pt,a4]{article}
\input{colordvi}
\begin{document}

\title{\bf Structured Adiabatic Quantum Search}

\maketitle
\begin{center}
Daria Ahrensmeier, Randy Kobes, Gabor Kunstatter, Haitham Zaraket\\[0.3cm]

{\it Physics Dept and Winnipeg Institute for Theoretical Physics\\
The University of Winnipeg,\\
515 Portage Avenue,
Winnipeg, Manitoba R3B 2E9, Canada,}
\end{center}
\vspace*{0.3cm}
\begin{center}
Saurya Das\\[0.3cm]
{\it Dept of Math and Statistics \\
University  of New Brunswick, Canada.}
\end{center}

\begin{abstract}
We examine the use of adiabatic quantum algorithms to solve structured, or nested, search problems. We construct suitable time dependent Hamiltonians and
derive the computation times for a general class of nested searches involving
$n$ qubits. As expected, we find that as additional structure is included, the Hamiltonians become more
local and the computation times decrease.
\end{abstract}
\section{Introduction}
 Adiabatic Quantum Computation (AQC) is a relatively new paradigm
in the field of quantum computing. Whereas in the standard model of QC (SQC), an
algorithm is defined as a sequence of discrete unitary transformations, AQC
(see, e.g., \cite{Farhi:0001,Roland:0107,Childs:0108,Das:0111,Smel:0202,Dam:0206})
considers the continuous time evolution of the quantum system,
described by the Schr\"{o}dinger equation
\begin{equation}
 i \frac{d}{dt} |\Psi (t)\rangle = H(t) | \Psi(t)\rangle \, 
\end{equation} 
with a time dependent Hamiltonian $H(t)$. 
A computational problem which has been studied extensively in both SQC and AQC is the 
unstructured search. In its simplest form the unstructured search
corresponds to finding a single marked state (needle) in a completely unstructured
database (haystack) of $N$ states. In the case that the physical system consists of 
$n$ two state particles (qubits), $N=2^n$ corresponds to the size of the
complete Hilbert space for the system. Classically a random search of $N$ objects 
requires on average $N$ operations to pick out the marked object. It was shown by 
BBBV \cite{BBBV} that for SQC, the lower bound on the corresponding number of steps is
 of order $\sqrt{N}$. Grover was the first to construct a
specific QC algorithm that achieved this lower bound \cite{Grover:97}.

In the context of AQC, the physical quantity that seems to most closely correspond to
the number of steps of a SQC algorithm is the time required to do the computation. It is therefore not surprising that under fairly standard assumptions, it can be shown that the time required for the unstructured search in AQC increases
 as $\sqrt{N}$ \footnote{In a dynamical
quantum system one can always shorten the time by increasing the energy.
The consequences for AQC of increasing the energy temporarily have recently been
 analyzed in \cite{Das:0204}.}.
 
In many physical systems, it is possible to use additional structure in order to shorten the search. For example, consider a database of items with  $n$ distinguishable
 bits. A maximally structured classical search would check one bit at a time
in order to find the correct marked state. Such a search would require 
$O(n)$ operations. The key is that, by using the existing structure, one is able
in principle to search $n$ two-dimensional objects, instead of a random search
among $N=2^n$ objects.  

The above example is a special case of a general class of nested searches, which have  recently been analyzed \cite{Cerf:00} 
in the context of SQC.
The purpose of the present paper is to study  structured {\em adiabatic} quantum search.
We find that additional structure shortens the running time and makes the Hamiltonian
more local (concerning the interactions of the system, as is explained below). 
Our results for the running time 
are  consistent with the conjecture (cf. \cite{Cerf:00}) that 
(adiabatic) quantum computation improves the running time of the corresponding 
classical algorithm by a square root; due to the fact that quantum computation 
manipulates probability amplitudes, whose squares give the corresponding probabilities.

 We start with a short description
of AQC in general. In Section 3 we
show that the unstructured search studied by Grover leads to a spatially non-local Hamiltonian when
 considered
in the conventional model for the implementation of quantum algorithms, namely 
two-level
quantum systems (qubits), such as spin-1/2 particles. That is, the unstructured adiabatic search requires an $n$-body
interaction for an $n$-qubit system. In Section 4, we
define  the notion of a structured adiabatic quantum search Hamiltonian and show that it is in general local, in the sense referred to above: it couples only
a subset of the total number of qubits.
We then start our detailed analysis of nested searches with the most intuitive case, the maximally structured search, and consider
its adiabatic evolution and running time, which is shown to scale as $T=O(\sqrt{n})$.
Finally, we study the general case and obtain its running time, comparing different ways of
structuring the system. 
\section{Adiabatic Quantum Computation}
The method of AQC is based on the use of the adiabatic theorem for considering quantum
 computing as continuous time evolution from some easily prepared initial
state $|\psi_i\rangle$ to a final state $|\psi_f\rangle$ that encodes the solution to
the computational problem.
Specifically, the  adiabatic theorem (see, e.g. \cite{Bransden}) states the following: a system which is 
described by a  time-dependent Hamiltonian $H(t)$
will stay close to the instantaneous ground state of $H(t)$ provided that the 
time evolution is slow enough. In quantitative terms, 
after time $T$
\begin{equation}
 |\langle E_0;T|\Psi(T)\rangle |^2 \geq 1-\epsilon ^2\, ,
\end{equation}
where $E_0$ is the lowest energy eigenvalue and $\epsilon \ll 1$,
provided that:
\begin{equation}
 \frac{|\langle E_1;t |\frac{dH}{dt}| E_0;t \rangle |}{\omega_{min}^2}\leq \epsilon\, 
\label{adiabat1}
\end{equation}
where 
\begin{equation}
 \omega_{min}=\min_{0\leq t \leq T} [E_1(t) -E_0 (t)]
\end{equation}
is the minimum gap between the lowest two energy eigenvalues $E_0$ and $E_1$.
Thus, if (\ref{adiabat1}) is satisfied, a measurement after time $T$ gives as result the solution with almost certainty,
i.e. with probability $\approx 1-\epsilon^2$. 

The time-dependent Hamiltonian that is traditionally used in AQC is constructed as the linear combination
\begin{equation}
 H(t)=f(t)\,H_i + g(t)\,H_f
\label{adiab_ham}
\end{equation}
where $|\psi_i\rangle$, $|\psi_f\rangle$ are the ground states of $H_i$, $H_f$,
 respectively and
 $f(t)$ and $g(t)$ are usually considered to be monotonic functions
such that $f(0)=1$, $f(T)=0$, $g(0)=0$ and $g(T)=1$, where $T$ is the total computation time. 

It will be useful for what follows to note that the adiabatic theorem (\ref{adiabat1})
 can be modified \cite{ADKKZ}
 for the case of an
$m$-fold
degenerate first excited state. In this case, it reads
\begin{equation}\label{DADT}
 \sum_{i=1}^m \frac{|\langle E_1 |\frac{dH}{dt}|E_0\rangle _i |^2}{\omega_{min}^4}
 \leq \epsilon^2\, .
\end{equation}
\section{Unstructured search in AQC}
The goal of the search algorithm is to find a marked object $|m\rangle$ in an
unstructured database of size $N$ in as few steps as possible. Classically, one has
to look at $O(N)$ objects to find the marked one. One advantage of the 
quantum search is that it manipulates the amplitude of the marked state, leading to
a quadratic amplification of the probability. The initial state is the superposition of
all states with equal weight,
\begin{equation}
 |\Psi_0\rangle\equiv|\Psi(0)\rangle = \sum _{i=1}^N \frac{1}{\sqrt{N}} |i\rangle\, .
\end{equation}
A natural choice for a  Hamiltonian
that has this state as its ground state is:
\begin{equation}
 H_i = 1 - | \Psi_0\rangle \langle \Psi_0 | \, .
\end{equation}
Correspondingly one must choose (cf. Section 4.1),
\begin{equation}\label{Hf}
 H_f = 1 - |m\rangle \langle m| \, ,
\end{equation} 
which has the ground state $|m\rangle$.
 $H_f$ is in a sense the adiabatic analogue of the ``oracle'' 
$\hat{I}-2|m\rangle\langle m|$ used
in the Grover search algorithm. It projects out the marked state $|m\rangle$. 
Thus, in the framework of AQC, the initial state $|\Psi_0\rangle$ evolves into 
$|m\rangle$ in time $T$, given that the Hamiltonian (\ref{adiab_ham})
varies slowly enough with time. In the simplest case \cite{Farhi:0001} $f(t)=1-t/T$
and $g(t)=t/T$, the minimum running 
time increases with $N$ at the same rate as for the classical search: $T=O(N)$.
However, one can improve this by choosing $f(t)$ and $g(t)$ to vary most rapidly when the gap $\omega$ is the largest. For example, one can use the
approach of \cite{Roland:0107}, in which $f(t)=1-s(t)$, $g(t)=s(t)$ with $s(0)=0$
and $s(T)=1$. This choice yields an adiabaticity condition of the form:
\begin{equation}
 \frac{|\langle E_1;s|\frac{dH(s)}{ds}|E_0;s\rangle|}{\omega^2(s)}
 \left|\frac{ds}{dt}\right| \leq \epsilon\, .
\label{bound1}
\end{equation}
One is then free to choose $s(t)$ so that the bound in (\ref{bound1}) is 
saturated for all $t$. This yields  
a running time  $T=O(\sqrt{N})$. Recently, it was shown \cite{Das:0204} that
with a more general choice of the functions $f$ and $g$  in the construction of $H(t)$
 the running time can -- in principle -- be made independent 
of $N$, but at the cost of requiring a large amount of energy to
 be {\em temporarily} 
injected into the system. 

We are now lead  to the question of how to implement a generic AQC 
algorithm in a real quantum
computer, and what kind of physical systems might be suitable.
The conventional scheme for the implementation  of quantum computation 
considers
 a system of uncorrelated 2-level
quantum systems (qubits), e.g. spin $1/2$-particles. The 1-qubit space $\mathcal{H}_i$
is spanned by $|0\rangle =  {1\choose 0}$ and $|1\rangle =  {0\choose 1}$.
The n-qubit space is given by the tensor product
\begin{equation}
 \mathcal{H}=\mathcal{H}_1\otimes\mathcal{H}_2\otimes ... \otimes\mathcal{H}_n
\end{equation}
with the basis $\{|\alpha_1, ..., \alpha_n\rangle =|\alpha_1\rangle\otimes ...
 \otimes |\alpha_n\rangle | \alpha_i\in\{0,1\}\}$. The marked state is
$|m\rangle = |z_1 ... z_n\rangle = |z_1\rangle\otimes ... \otimes |z_n\rangle$
with $z_i\in \{0,1\}$. In this framework, the final Hamiltonian 
\begin{equation}
 H_f = 1 -  |m\rangle\langle m|
\label{hf}
\end{equation}
can be written in terms of Pauli spin matrices
\begin{eqnarray}
 H_f & = & 1-\prod_{i=1}^n \left(\frac{1}{2}
              (1-(-1)^{z_i}\sigma_z^{(i)})\right)\\
     & \equiv & \sum_{i=0}^n c_i \prod_{j=0}^i \sigma_{z}^{(j)}\, .
\end{eqnarray}
In this form it is clear that the Hamiltonian contains a product of $n$ spin matrices,
 which implies interactions
among all qubits. This non-locality may put constraints on possible implementations of such algorithms,
at least for spin systems\footnote{If one considers AQC only as a possible simulation 
of SQC, the non-locality is irrelevant.}.

In the next section, we describe how, by using additional structure, the search can be modified to reduce the
spatial non-locality in the Hamiltonian, while keeping the initial and final ground state the same. This structured search also provides a shorter computation time than the unstructured one.

\section{Structured adiabatic search} 
The adiabatic theorem requires the evolution of the quantum system to be slow enough to
keep the system in its ground state for all times. It seems plausible that it is
``easier'' for the system to stay in its ground state if there are no long range
 interactions,
but only  2 or 3-qubit interactions. In order to find a Hamiltonian that realizes this
idea, it is useful to consider the problem in
 the language of decision clauses etc.(cf. \cite{Farhi:0001}):
\subsection{Satisfiability problems, clauses and Hamiltonians} 
Search algorithms belong to the class of satisfiability problems. A satisfiability
problem is a combination of decision clauses $C_i$ which have an
element of $\{0,1\}$ (true or false) as output which depends on the values of some set 
of the qubits of the system. A formula is an $n$-bit instance of satisfiability,
\begin{equation}\label{formula}
 F \equiv C_1\wedge C_2\wedge ...\wedge C_m\, ,
\end{equation}
i.e. in general $m$ clauses acting on $n$ bits.
The action of a clause or formula on an assignment can be translated into the action of
a Hamiltonian on the states of a physical system (up to an overall factor): 
Formula (\ref{formula}) is mapped into
\begin{equation}
 H(t) = H_1(t) + H_2(t) + ... + H_n(t)\, ,
\end{equation}
where each $H_i$ is constructed from the clause $C_i$ and acts only on the bits of this
clause. For example, if $C_i$ is a two-bit clause, $H_i$ contains only operators that
act on these two bits, it can have at most a two-body interaction term. The action of
$H_i$ on the bits absent in $C_i$ is given by the identity operator. The initial state 
is the ground state of $H(0)$, and the ground state of $H(t=T)$ satisfies all
the clauses in the formula.

In this language, the unstructured search algorithm is a single clause that acts on
 $n$ qubits
and  has a unique (but unknown) satisfying assignment, $|m\rangle$.

It should be noted that the form of $H_f$ given in (\ref{Hf}) is the most general one available for an oracle search. By definition, an
oracle clause acts as
\begin{eqnarray}
 C_m(|m\rangle) & = & 0\\
 C_m(|i\rangle) & = & 1 \;\;\forall i\neq m\, .
\end{eqnarray}
The corresponding adiabatic Hamiltonian must therefore  fulfill 
\begin{eqnarray}
 H(t=T)|m\rangle & = & H_f|m\rangle  =  0\\
 H(t=T)|i\rangle & = & H_f|i\rangle  = f(T)|i\rangle\;\; \forall i\neq m\, ,
\end{eqnarray}
where $f(T)$ should be the same for all $i \neq m$. Therefore, $H_f$ is diagonal,
and it reads
\begin{eqnarray}
 H_f & = & f(T)\sum_{i\neq m} |i\rangle\langle i|\nonumber \\
     & = & f(T)[1 -|m\rangle\langle m|]\,.
\end{eqnarray}
\subsection{Maximally structured search} 
The unstructured search, which is non-local in space, corresponds to a single
$n$-bit clause.
In order to find a local Hamiltonian, it is therefore the most intuitive alternative 
to consider a formula of $n$ 1-bit clauses,
\begin{equation}
 F= C_1 \wedge C_2\wedge ... \wedge C_n,
\end{equation}
where $C_i$ is satisfied if and only if the i-th bit has the required value $z_i$ for
the marked state $|m\rangle =|z_1 z_2 ... z_n\rangle$. We call this ``maximally
structured'' because it corresponds to the maximal splitting of the
system:  The adiabatic Hamiltonian
which has as ground state the unique satisfying assignment of F is
a sum of 1-bit versions of the oracle Hamiltonian,
\begin{equation}
 H_f^{str} = \sum_{i=1}^n h_i
\end{equation}
with
\begin{eqnarray}
 h_i & = & 1\otimes  ... \otimes 1 \otimes(1-|z_i\rangle\langle z_i |)
            \otimes 1 \otimes ... \otimes 1\nonumber\\
     & = & 1\otimes  ... \otimes 1 \otimes
          \left[\frac{1}{2}(1-(-1)^{z_i}\sigma_z)\right]
           \otimes 1 \otimes ... \otimes 1\nonumber\\
     & \equiv & \frac{1}{2} (1 - ( -1)^{z_i} \sigma_z ^{(i)})\, ,
\end{eqnarray}
which act as oracle on the i-th bit and as identity on the others. The application of
$H_f^{str}$ on a general state $|\alpha_1 \alpha_2 ... \alpha_n\rangle$ gives the 
number of qubits
with $\alpha_i\neq z_i$. This illustrates the crucial difference between
this procedure and  the unstructured search. 
The latter leads to the same answer, $1$, for all unsatisfying assignments, whereas in the present case the 
query gives the number $F[|\alpha\rangle]$ of unsatisfied clauses.
 Thus, one gets  additional
information. 

As before, the initial state $|\Psi_0\rangle$ is the ground state of
$H_i = 1-|\Psi_0\rangle\langle\Psi_0|$, which can be written as
\begin{equation}
 H_i^{str}\equiv \frac{1}{2} \sum_{i=1}^n (1-\sigma_x^{(i)})\, .
\end{equation}
After some algebra (cf. \cite{ADKKZ}), the eigenvalues of the Hamiltonian
\begin{equation}
 H^{str}(t)= \frac{1}{2}f(s)\sum_{i=1}^n (1-\sigma_x^{(i)}) +
             \frac{1}{2}g(s)\sum_{i=1}^n (1-(-1)^{z_i}\sigma_z^{(i)})
\end{equation}
 are found to be
\begin{equation}
 E_m^{str}=\frac{n}{2}(f+g)-m\sqrt{f^2+g^2}
\end{equation}
where $m=\sum m_i$, with $m_i=\pm 1/2$ representing the spin up and down of the i-th 
particle, corresponding to $\alpha_i = 0,1$. The ground state is obtained for
$m_i =1/2\;\forall\, i$, i.e. $m=n/2$.
The first excited state ($m=n/2 -1$) is $n$-fold degenerate.
The crucial condition for the adiabatic theorem to hold is  a non-vanishing gap
 between the ground state and the first excited state, which is fulfilled:
\begin{equation}
\omega\equiv E^{str}_{1}-E^{str}_0=\sqrt{f^2+g^2}\neq 0\, .
\end{equation}
Note that this result corresponds to Eq.(14) of \cite{Das:0204} for $2^n=N=2$, i.e.
for $n=1$, the single qubit adiabatic search.
The running time is determined by the adiabatic theorem, which in the case of a degenerate
first excited state leads to the condition (\ref{DADT}).
The transition probability to
each of the first excited states is proportional to the square of  \\
\begin{equation}\label{prob}
_i\langle E^{str}_1|\frac{dH}{dt}| E^{str}_0 \rangle=-\frac{(-1)^{z_i}}{2}
 \frac{\dot{f}g-\dot{g}f}{\sqrt{f^2+g^2}}\; ,
\end{equation}
so the total transition probability is $n$ times Eq.(\ref{prob}) squared
(cf. Eq.(15) in \cite{Das:0204} for $N=2$).
For an estimate of the running time, we consider the simplest case in which
 $f$ and $g$ are  linear in $s$,
$f(s)=1-s(t)$ and $g(s)=s(t)$,  
which, after optimizing $s(t)$ as in the previous section, results in (cf.\cite{ADKKZ})
\begin{equation}
 T = \sqrt{n}/\epsilon = \sqrt{\log N}/\epsilon\; ,
\end{equation}
 polynomial in $n$. It should be noted  that a similar polynomial time algorithm has
 been studied in \cite{Farhi:0001}.
\subsection{General case of  structured search} 
The transformation of the $n$-bit clause into $n$ 1-bit clauses is not the only 
possibility
of introducing structure to the search. For example, one could also consider clauses 
acting on 2 qubits,
\begin{equation}
 F_2=C_{12}\wedge C_{34}\wedge ... \wedge C_{(n-1)n}
\end{equation}
corresponding to the Hamiltonian
\begin{equation}
 H_f^{str2} =\sum_{i=0}^{n/2-1}h_{[2i+1][2(i+1)]}\,,
\end{equation}
where 
\begin{equation}
 h_{[2i+1][2(i+1)]} =1 \otimes  ... \otimes 1\otimes 
             (1-|z_{2i+1} z_{2(i+1)}\rangle \langle z_{2i+1} z_{2(i+1)}|)
               \otimes 1\otimes ... \otimes 1
\end{equation}
acts on the two neighboring qubits as oracle and as identity on the others.

A more interesting case with respect to realization would be 
\begin{equation}
 H_f=\sum_{i=1}^{n-1} h_{i[i+1]}\, ,
\end{equation}
with overlap of the interacting qubit-pairs (see the discussion in \cite{ADKKZ}).

In general, any splitting of the $N=2^n$-dimensional Hilbert space into $m$ smaller 
spaces
of dimensions $N_i=2^{n_i}$ with $\sum n_i =n$ is possible. In order to compare the 
effects of different ways of splitting, we calculate the running time for several 
cases numerically (cf. \cite{Das:0204} for the method). 

\begin{table}
\begin{center}
\begin{tabular}{|c|c|c|c|c|}
$m$  &  $n/m$ &  $\epsilon T$  &  $\alpha$ & $\beta$ \\
\hline 
1  &     6   &    7.94  &   0.9962  &  $\infty$ \\
2   &    3   &    3.74  &   0.9518   & 3.8074  \\
3  &     2   &    3.00  &   0.8842   & 2.0000  \\
6  &     1    &   2.45  &   0.7211   & 1.0000 \\
\hline
\end{tabular}
\end{center}
\caption{Comparison of times for various splittings for $n=6$.}
\end{table}

\begin{table}
\begin{center}
\begin{tabular}{|c|c|c|c|c|}
$m$  &  $n/m$ &  $\epsilon T$  &  $\alpha$ & $\beta$ \\
\hline 
1    &     30   &    32768.00 &    1.0000 &   $\infty$ \\
2     &    15   &    256.00   &  1.0000 &   16.0000\\
3    &     10   &    55.40  &   0.9999 &   7.3084\\
5    &     6    &   17.75   &  0.9973 &   3.5743\\
6    &     5    &   13.64   &  0.9940 &   2.9165\\
10    &     3   &    8.37  &   0.9695  &  1.8451\\
15    &     2  &     6.71  &   0.9297 &   1.4057\\
30   &      1   &    5.48  &   0.8307 &   1.0000\\
\hline
\end{tabular}
\end{center}
\caption{Comparison of times for various splittings for $n=30$.}
\end{table}
\par

In order to investigate the scaling of $T$ with the splitting, we define
the two coefficients $\alpha,\beta$ implicitly by:
\begin{eqnarray}
 \epsilon T & = & (\sqrt{m}\sqrt{2^{n/m}})^{\alpha}\nonumber\\
 \epsilon T & = & (\sqrt{m})^{\beta}\,.
\end{eqnarray}

We find (see Tables 1 and 2)
that for no splitting ($m=1$), $\alpha \to 1$ for large $n$, which means the running time
 is $T=O(\sqrt{N})$, as is known from the unstructured search.

 For  $m=2$, i.e.
$n=n_1 +n_2$, the time is shortened, and the optimal value in this case 
is achieved for $n_1 =n_2$.

For $m\geq 2$, the running time is even shorter, and it scales with the root of the
 dimension of the largest
Hilbert space of the splitting, so it is again optimal for equal values of $n_i$ for all $i$. This can also 
be seen from the expression for the running time,
\begin{equation}\label{time}
 T= \frac{1}{\epsilon}\int_0^1 \, ds\, |\dot{f}g-\dot{g}f|
    \sqrt{\sum_{i=1}^m \frac{N_i -1}{N_i^2}\frac{1}{\omega_i^6}}
\end{equation}
with 
\begin{equation}
 \omega_i=\sqrt{(f-g)^2 +\frac{4}{N_i}fg}\, 
\end{equation}
the gap for the $i$-th subsystem.
 For splitting into $m$ parts of equal size,
 $N_i = 2^{n/m}\;\forall\,i$, and Eq.(\ref{time}) simplifies to
\begin{equation}
 T=\frac{1}{\epsilon}\sqrt{m}\frac{\sqrt{N_i-1}}{N_i}\int_0^1 \, ds\, 
    \frac{|\dot{f}g-\dot{g}f|}{[(f-g)^2+\frac{4}{N_i}fg]^{3/2}}
\end{equation}
(which reduces to Eq.(20) in \cite{Das:0204} for $m=1$) where
 the integral scales with $N_i$.
 For $n/m \gg 1$, we find $\alpha \sim 1$, and
\begin{equation}
 \epsilon T=\sqrt{m}\; \sqrt{2^{n/m}}\, . 
\end{equation}

For maximal splitting, $m=n$ and $n_i=1\,\forall i$, leading to $\beta = 1$,
which results in
$T=O(\sqrt{n})$, as shown above.

\section{Conclusions} 
We have shown how additional structure affects both the running time and locality of the Hamiltonian required for an adiabatic quantum search algorithm. As expected, the more structure, the more local the Hamiltonian and the shorter the running time. In fact, the expressions we have obtained for the running time of the general structured search suggest
 strongly that adiabatic quantum computation consistently improves the running time for the corresponding classical algorithm by a square root. This supports the  conjecture 
(cf. \cite{Cerf:00}) that the speed-up achieved by quantum 
computation can be directly attributed to the fact that the quantum computation
 algorithm manipulates probability amplitudes, whose squares give the corresponding probabilities. It would be interesting to see whether this conjecture can be
proven in more general contexts.
  Another interesting topic for future work concerns what happens in the more general case that the pairwise interactions overlap. These and other related issues will be addressed elsewhere.

\section*{Acknowledgments}
It is a pleasure to thank N.J. Cerf, J. Currie, E. Farhi, S. Gutmann, S. Hamieh, 
and J. Roland  for helpful discussions at various stages of this work.


\begin{thebibliography}{99}
\bibitem{Farhi:0001} 
         E. Farhi, J. Goldstone, S. Gutmann, and  M. Sipser,
         {\em Quantum Computation by Adiabatic Evolution},
         quant-ph/0001106.
\bibitem{Roland:0107} 
         J. Roland and N.J. Cerf,
         {\em Quantum Search by Local Adiabatic Evolution},
          Phys. Rev. A 65, 042308; quant-ph/0107015. 
\bibitem{Childs:0108}
         A.M. Childs, E. Farhi, and J. Preskill,
         {\em Robustness of adiabatic quantum computation},
         quant-ph/0108048.
\bibitem{Das:0111}
         S. Das, R. Kobes, and G. Kunstatter,
         {\em Adiabatic Quantum Computation and Deutsch's Algorithm},
         Phys. Rev. {\bf A 65} (2002) 062310;
         quant-ph/0111032.
\bibitem{Smel:0202}
         V.N. Smelyanskiy and U.V. Toussaint,
         {\em Number Partitioning via quantum adiabatic computation},
         quant-ph/0202155.
\bibitem{Dam:0206}
         W. van Dam, M. Mosca, and U. Vazirani,
         {\em How powerful is Adiabatic Quantum Computation?},
         Proceedings of the 42nd Annual Symposium on Foundations of
         Computer Science, pp. 279-287 (2001);
         quant-ph/0206003.
\bibitem{BBBV}
        C. Bennett, E. Bernstein, G. Brassard, U. Vazirani,
        {\em Strengths and weaknesses of quantum computing}, 
        SIAM Journal on Computing, Volume {\bf 26}, No. 5 (1997) 1510;
        quant-ph/9701001.
\bibitem{Grover:97}
         L. K. Grover, 
         {\em Quantum Mechanics helps in searching for a needle in a haystack},
         Phys. Rev. Lett. {\bf 79} (1997) 325; quant-ph/9706033.
\bibitem{Cerf:00}
         N.J. Cerf, L.K. Grover, and C.P. Williams,
         {\em Nested quantum search and structured problems},
         Phys. Rev. A 61, 032303.
\bibitem{Bransden}
         A. Messiah, {\em Quantum Mechanics} Vol.II, Amsterdam: North Holland,
         New York: Wiley (1976);
         B.H. Bransden, C.J. Joachain, {\em Quantum Mechanics}, Pearson Education
         (2000).
\bibitem{Das:0204}
        S. Das, R. Kobes, and G. Kunstatter,
        {\em Can the Adiabatic Quantum Search Algorithm be Turbo-Charged?},
         quant-ph/0204044.
\bibitem{ADKKZ}
        D. Ahrensmeier, S. Das, R. Kobes, G. Kunstatter, and H. Zaraket,
        in preparation.
\end{thebibliography}
\end{document}